\documentstyle[epsfig]{mn}
\newif\ifAMStwofonts
\AMStwofontstrue                % <- comment this out if AMS fonts not available
\DeclareMathVersion{bold}                    % <- seems to be needed with mathptm
\DeclareMathAlphabet{\mathbfit}{OT1}{cmr}{bx}{it}
\SetMathAlphabet\mathbfit{bold}{OT1}{cmr}{bx}{it}
\DeclareMathAlphabet{\mathbfss}{OT1}{cmss}{bx}{n}
\SetMathAlphabet\mathbfss{bold}{OT1}{cmss}{bx}{n}
\ifAMStwofonts
  \ifCUPmtlplainloaded \else
    \DeclareSymbolFont{UPM}{U}{eur}{m}{n}
    \SetSymbolFont{UPM}{bold}{U}{eur}{b}{n}
    \DeclareSymbolFont{AMSa}{U}{msa}{m}{n}
    \DeclareMathSymbol{\upi}{0}{UPM}{"19}
    \DeclareMathSymbol{\umu}{0}{UPM}{"16}
    \DeclareMathSymbol{\upartial}{0}{UPM}{"40}
    \DeclareMathSymbol{\leqslant}{3}{AMSa}{"36}
    \DeclareMathSymbol{\geqslant}{3}{AMSa}{"3E}

     \let\le=\leqslant

  \fi
\fi
\title[Combined MEM and MHW analysis of CMB observations]
{Combining maximum-entropy and the mexican hat wavelet to
reconstruct the microwave sky}
\author[Vielva et al.]{P. Vielva$^{1,2}$, R. B. Barreiro$^3$,
M. P. Hobson$^3$, E. Mart{\'\i}nez-Gonz{\'a}lez$^1$, A. N. Lasenby$^3$, 
\and J. L. Sanz$^1$ and L. Toffolatti$^4$ \\
$^1$ Instituto de F{\'\i}sica de Cantabria (CSIC -- UC), Fac. Ciencias, Avda. de los
Castros s/n, 39005 Santander, Spain\\
$^2$ Departamento de F{\'\i}sica Moderna, Universidad de Cantabria, 
Avda. de los Castros s/n, 39005 Santander, Spain\\
$^3$ Astrophysics Group, Cavendish Laboratory, Madingley Road,
Cambridge CB3 0HE, UK\\
$^4$ Departamento de F{\'\i}sica, Universidad de Oviedo, c/ Calvo Sotelo
s/n, 33007 Oviedo, Spain}
\date{\today}
\begin{document}
\maketitle

\begin{abstract}

We present a maximum--entropy method (MEM) and
`Mexican Hat' wavelet (MHW) joint analysis to recover the
different components of the microwave sky from simulated observations
by the ESA Planck satellite in a small patch of the sky ($12.8\times 12.8$ deg$^2$).
This combined method allows one to improve the CMB, Sunyaev--Zel'dovich
and Galactic foregrounds separation achieved by the MEM technique alone.
In particular, the reconstructed CMB map is free from any bright
point source contamination.
The joint analysis also produces point source catalogues at each
Planck frequency which are
more complete and accurate than
those obtained by each method on its own. 
The results are especially improved at high frequencies where infrared
galaxies dominate the point source contribution.
Although this joint technique has been performed on
simulated Planck data, it could be easily applied to
other multifrequency CMB experiments, such as the
forthcoming NASA MAP satellite or the recently performed
Boomerang and MAXIMA experiments.
\end{abstract}

\begin{keywords}
methods: data analysis -- techniques: image processing -- cosmic
microwave background
\end{keywords}

\section{Introduction}
The cosmic microwave background (CMB) is one of the 
most powerful observational tools for understanding our Universe.
Indeed, an accurate knowledge of the CMB anisotropies
can place tight constraints on fundamental parameters such as
the amount of matter in the Universe and its overall
geometry. Observations of the CMB also
allow one to discriminate between different
models of structure formation.

Recent CMB experiments such as Boomerang (de Bernardis et al. 2000)
and MAXIMA (Balbi et al. 2000) have set strong constraints on 
the geometry of the Universe, showing that it is close to spatially flat.
Nevertheless, there remain numerous unbroken degeneracies in
the full set of parameters that define the currently favoured 
inflationary CDM cosmological
model. In order to resolve these degeneracies a new generation of 
CMB satellite experiments is currently in preparation, most notably
the NASA MAP mission (Bennet et al. 1996) and the ESA Planck Surveyor
(Puget et al. 1998, Mandolesi et al. 1998) . These
experiments will provide high-resolution all-sky maps
of the CMB anisotropies which will allow a highly accurate estimation
of a large set of cosmological parameters.
%with sufficient accuracy to allow the
%unambiguous estimation of the full set of cosmological parameters.

An important issue for CMB satellite missions is the 
separation of foreground emission from the
CMB signal. An accurate means of performing this separation
is vital in order to make full use of the high-resolution CMB maps
these experiments will produce. The main foregrounds
to be separated from the CMB signal are those
due to our own Galaxy (dust, free-free \& synchrotron emission)
and extragalactic emission due, principally,
to the Sunyaev--Zel'dovich effect (both thermal and kinetic)
and point sources.

A preliminary application of neural-network techniques in this field
has recently been performed with promising results (Baccigalupi et
al. 2000), but this
approach is not at present sufficiently sophisticated 
to accommodate multifrequency data arising from convolutions with 
beams of different sizes and subject to different
levels of instrumental noise.
Nevertheless, more traditional techniques 
based on the maximum-entropy method (MEM)
have been shown to provide an
efficient and accurate way of performing the component separation
(Hobson et al. 1998; H98, hereafter).

As one might expect, the MEM technique is particularly successful at
using multifrequency data to identify foreground emission from
physical components whose spectral signatures are (reasonably) well-known.
It is therefore not surprising that the most problematic foreground
to remove is that due to extragalactic point sources. These differ
from the other components in the important respect that
the spectral behaviour of point sources differs from one source to the
next and, moreover, is notoriously difficult to predict.
We can, however, make some headway by at least identifying the likely
populations of point sources at different observing frequencies.
Our knowledge of point source
populations is increasing with new observations,
but there are still great uncertainties. In the absence of extensive
observations at microwave frequencies, the currently-favoured approach
is to use the Toffolatti et al.
(1998) model, which provides an estimate of point source populations
based on existing observations and basic physical mechanisms.
This model can be used to generate simulated point source emission at 
different observing frequencies.
From 30 to 100 GHz, the main point source emission is due
to radio selected flat--spectrum AGNs (radio-loud quasars, blazars,
etc.). From 300 to 900 GHz, infrared selected sources
- starburst and late type galaxies at intermediate to low
redshift and high redshift ellipticals - account
for most  point source emission. At intermediate frequencies,
both populations contribute approximately equally.

The problem of removing emission due to point sources from CMB
observations has been addressed by several authors.
For example, Tegmark \& Oliveira-Costa (1998; TOC98, hereafter)
suggest a straightforward
harmonic filter technique that is optimised to detect and 
subtract point sources, whereas Hobson et al. (1999; H99, hereafter)
propose treating point source emission as an additional generalised `noise'
contribution within the framework of the MEM approach discussed in H98.
Tenorio et al. (1998) present a wavelet technique
to subtract point sources, but the wavelet basis used in this work is not
the optimal one for this purpose. This point is addressed in
Sanz, Herranz \& Mart{\'\i}nez-Gonz{\'a}lez (2000), 
where it is shown that the `Mexican Hat' wavelet (MHW) is in fact optimal
for detecting point sources under reasonable conditions,
the most important assumption being that the beam is well approximated
by a Gaussian.

The application of this wavelet to realistic
simulations has been presented in Cay{\'o}n et al. (2000; C00, hereafter) and
extended in Vielva et al. (2001; V01, hereafter) . The main advantage of
the MHW method over the previous works is that
the algorithm does not require any assumptions to be made regarding the
statistical properties of the point source population or the
underlying emission from the CMB (or other foreground components).

The aim of this paper is to show how the MEM and MHW approaches
are in fact complementary and
can be combined to improve the accuracy of the separation of
diffuse foregrounds from the CMB and increase the number of points
sources that are identified and successfully subtracted.
The technique proposed in this paper is as follows. First, we apply the
MHW to the multifrequency simulated Planck maps to detect
the brightest point sources at each observing frequency, which are
subtracted from the original data. The MEM
algorithm is then applied to these processed maps in order to recover
the rest of the components of the microwave sky. These reconstructed
components are then used as inputs to produce `mock' data and subtracted
from the original data maps. Since we expect our reconstructions to be
reasonably accurate, the residuals maps obtained in this way
would mostly contain noise and the contribution
from point sources. Finally, we apply the MHW on these residuals maps in
order to recover a more complete and accurate point source catalogue.

In this paper the combined method is applied to simulated observation
by the Planck satellite, but the technique could easily be applied to
MAP data or to existing multifrequency observations by the Boomerang
or MAXIMA balloon experiments. The paper is organized
as follows. In the next section we present a brief
overview of the MEM and MHW techniques, outlining in particular
the advantages and shortcomings of each approach.
We then explain why the two approaches can be successfully combined to
produce a more powerful joint analysis scheme.
In section~3 we summarise how the simulated Planck observations were
performed. In section~4 we present the results of a component
separation based on the new joint analysis method, and discuss the
improved accuracy of the reconstructions of the diffuse components.
In section~5 we concentrate on the recovery of the point
sources themselves and discuss the construction of point source
catalogues from Planck observations. Finally, our conclusions are
presented in section~6.

\section{The MEM and MHW techniques}

In this section, we briefly review the MEM and MHW techniques.
A complete description of the MEM component separation
algorithm can be found in H98. In addition, H99 describes how to
include point sources into the MEM formalism. We therefore provide
only a basic outline of the approach.
The MHW method is introduced in C00 and extended in V01, and so again
we give only a basic summary.

\subsection{The maximum-entropy method}

If we observe the microwave sky in a given direction ${\mathbfit x}$ at
$n_f$ different frequencies, we obtain an $n_f$-component data vector
that contains, for each frequency, 
the temperature fluctuations convolved with the beam in this direction
plus instrumental noise. The $\nu$th component of the data 
vector in the direction ${\mathbfit x}$ may be written as
\begin{equation}
d_{\nu}({\mathbfit x})=\int B_\nu(|{\mathbfit x}-{\mathbfit x}'|)
\sum_{p=1}^{n_c} F_{\nu p}\,s_p({\mathbfit x}')\,d^2{\mathbfit x}'
+ \xi_\nu({\mathbfit x}) 
+ \epsilon_\nu({\mathbfit x}).
\label{datadef}
\end{equation}
In this expression we distinguish between the contributions from
the point sources and the $n_c$ physical components 
for which it is assumed the
spectral behaviour is constant and reasonably well-defined (over the observed 
patch of sky). The latter are collected together in a signal vector with
$n_c$ components, such that $s_p({\mathbfit x})$ is the signal from the
$p$th physical component at some reference frequency $\nu_0$. 
The corresponding total emission at the observing frequency $\nu$
is then obtained by multiplying the signal vector by the $n_f \times
n_c$ frequency response matrix $F_{\nu p}$ that includes the spectral
behaviour of the considered components as well as the transmission of
the $\nu$th frequency channel. This contribution is then convolved
with the beam profile $B_\nu({\mathbfit x})$ of the relevant channel.
Since the individual spectral dependencies of the point sources are very
complicated, we cannot factorize their contribution in this way and
so they are added into the formalism as an extra `noise' term. Thus
$\xi_\nu$ is the contribution from point sources,
as observed by the instrument at the frequency $\nu$
(hence, convolved with the beam profile).
Finally, $\epsilon_\nu$ is
the expected level of instrumental noise in the $\nu$th frequency
channel and is assumed to be Gaussian and isotropic.

The assumption of a spatially-invariant beam
profile in (\ref{datadef}) allows us to perform the reconstruction
more effectively by working in Fourier space, since we may consider
each ${\mathbfit k}$-mode independently (see H98).  
Thus, in matrix notation, at each mode we have
\begin{equation}
\mathbfss{d} 
= \mathbfss{R} \mathbfss{s}+ \bxi +\bepsilon
= \mathbfss{R} \mathbfss{s}+ \bzeta,
\label{dataft2}
\end{equation}
where $\mathbfss{d}$, $\bxi$ and $\bepsilon$ are column vectors
each containing $n_f$ complex components and $\mathbfss{s}$ is a
column vector containing $n_c$ complex components. In the second
equality we have combined the instrumental noise vector $\bepsilon$
and the point-source contribution $\bxi$ into a single
`noise' vector $\bzeta$. The response matrix $\mathbfss{R}$ has
dimensions $n_f\times n_c$ and its elements are given by $R_{\nu
p}({\mathbfit k}) = \widetilde{B}_\nu({\mathbfit k})F_{\nu p}$. The aim of
any component separation/reconstruction algorithm is
to invert (\ref{dataft2}) in some sense, in order to obtain an
estimate $\hat{\mathbfss s}$ of the signal vector at each value of
${\mathbfit k}$ independently. Owing to the presence of noise, and the fact 
that the response matrix $\mathbfss{R}$ is not square and would, in
any case, have some small eigenvalues, a direct
inversion is not possible, and so some form of regularised inverse
must be sought. Typical methods include singular-valued decomposition,
Wiener filtering or the maximum-entropy method.

The elements of the signal vector $\mathbfss{s}$ at
each Fourier mode may well be correlated, this correlation being
described by the $n_c\times n_c$ signal covariance matrix
$\mathbfss{C}$ defined by
\begin{equation}
{\mathbfss C}({\mathbfit k}) = \langle 
{\mathbfss s}({\mathbfit k})
{\mathbfss s}^\dagger ({\mathbfit k})
\rangle,
\label{covdef}
\end{equation}
where the dagger denotes the Hermitian conjugate. 
Moreover, if prior information is available concerning these
correlations, we would wish to include it in our analysis. 
We therefore introduce the vector of `hidden' variables ${\mathbfss h}$,
related to the signal vector by
\begin{equation}
\mathbfss{s}=\mathbfss{L}\mathbfss{h},
\label{icfdef}
\end{equation}
where the $n_c \times n_c$ lower triangular matrix $\mathbfss{L}$ 
is obtained by performing a Cholesky decomposition
of the signal covariance matrix 
$\mathbfss{C}=\mathbfss{L}\mathbfss{L}^{\rm T}$. 
The reconstruction is then performed entirely in terms of
$\mathbfss{h}$ and the corresponding reconstructed signal vector is 
subsequently found using (\ref{icfdef}).

Using Bayes' theorem,
we choose our estimator $\hat{\mathbfss h}$ of the hidden vector
to be that which maximises the posterior probability given by
\begin{equation}
\Pr({\mathbfss h}|{\mathbfss d}) 
\propto \Pr({\mathbfss d}|{\mathbfss h})\Pr({\mathbfss h})
\label{bayes}
\end{equation}
where $\Pr({\mathbfss d}|{\mathbfss h})$ is the likelihood of obtaining
the data given a particular hidden vector and 
$\Pr({\mathbfss h})$ is the prior probability that codifies our
expectations about the hidden vector before acquiring any data.

As explained in H99, the form of the likelihood function 
in (\ref{bayes}) is given by
\begin{eqnarray}
\Pr({\mathbfss d}|{\mathbfss h}) 
& \propto & \exp \left(-\bzeta^\dag {\mathbfss N}^{-1} \bzeta \right)
\nonumber \\
& \propto & \exp \left[-({\mathbfss d}-{\mathbfss RLh})^\dag 
{\mathbfss N}^{-1} ({\mathbfss d}-{\mathbfss RLh})\right]
\label{likehd}
\end{eqnarray}
where in the last line we have used (\ref{dataft2}). 
The noise covariance matrix ${\mathbfss N}$ has 
dimensions $n_f \times n_f$ and at any given ${\mathbfit k}$-mode is given
by
\begin{equation}
{\mathbfss N}({\mathbfit k}) = \langle\bzeta({\mathbfit k})
\bzeta^\dagger ({\mathbfit k})\rangle.
\end{equation}
Therefore, at a given Fourier mode, the $\nu$th diagonal element of
${\mathbfss N}$ contains the ensembled-averaged power spectrum at that
mode of the instrumental noise plus the point source contribution to
the $\nu$th frequency channel. The off-diagonal terms give the
cross-correlations between different channels; if the noise is
uncorrelated between channels, only the point sources contribute to
the off-diagonal elements.

For the prior $\Pr({\mathbfss h})$ in (\ref{bayes}), we assume the
entropic form
\begin{equation}
\Pr({\mathbfss h}) \propto \exp[\alpha S({\mathbfss h},{\mathbfss m})]
\label{prior}
\end{equation}
where $S({\mathbfss h},{\mathbfss m})$ is the cross entropy of the 
complex vectors $\mathbfss{h}$ and $\mathbfss{m}$, where
${\mathbfss m}$ is a model vector to which ${\mathbfss h}$ defaults in 
absence of data. The form of the cross entropy for complex images 
and the Bayesian method for fixing the regularising parameter $\alpha$
are discussed in H98.  We note that, in 
the absence of non-Gaussian signals, the
entropic prior (\ref{prior}) tends to the Gaussian prior implicitly
assumed by Wiener filter separation algorithms, and so in this case
the two methods coincide. 

The argument of
the exponential in the likelihood function (\ref{likehd}) may 
be identified as (minus) the standard $\chi^2$
misfit statistic, so we may write $\Pr({\mathbfss d}|{\mathbfss h})
\propto \exp[-\chi^2({\mathbfss h})]$. Substituting this expression,
together with that for the prior probability given in (\ref{prior}),
into Bayes' theorem, we find that maximising the posterior probability
$\Pr({\mathbfss h}|{\mathbfss d})$ with respect to $\mathbfss{h}$ is
equivalent to minimising the function
\[
\Phi({\mathbfss h})=\chi^2({\mathbfss h}) 
- \alpha S({\mathbfss h},{\mathbfss m}).
\]
This minimisation can be performed using a variable metric minimiser
(Press et al. 1994) and requires only a few minutes of CPU time
on a Sparc Ultra workstation.

\subsection{The mexican hat wavelet method}

The MHW technique presented by C00 and V01 for identifying and
subtracting point sources operates on individual data maps.
Let us consider the two-dimensional data map $d_{\nu}({\mathbfit x})$ at
the frequency $\nu$. If the map contains point sources
at positions ${\mathbfit b}_i$ with fluxes or amplitudes
$A_i$, together with 
contributions from other physical components and instrumental noise,
then the data map is given by
\begin{equation}
d_\nu({\mathbfit x}) = \xi_\nu({\mathbfit x}) + n_\nu({\mathbfit x}) = 
\sum_i A_i B_\nu({\mathbfit x}-{\mathbfit b}_i) 
+ n_\nu({\mathbfit x}),
\end{equation}
where $B_\nu({\mathbfit x})$ is the beam at the observing frequency $\nu$,
and in this case the generalised `noise' 
$n_\nu({\mathbfit x})$ is defined as all 
contributions to the data map aside from the point sources.
%In the following analysis, it is assumed that 
%$n_\nu({\mathbfit x})$ is an homogeneous
%and isotropic random field with mean zero.

For the $i$th point source, we may define a `detection level'
\begin{equation}
D = \frac{A_i/\Omega}{\sigma_n},
\label{reald}
\end{equation}
where $\Omega$ is the area under the beam and $\sigma_n$ is the
dispersion of the generalised noise field $n_\nu({\mathbfit x})$.
In general, the detection level $D$ will be much less than unity
for all but the few brightest sources. This is the usual problem
one faces when attempting to identify point sources directly in the data
map.

As explained in C00, instead of attempting the
identification in real space, one can achieve better results by
first transforming to {\em wavelet space}. For a two-dimensional 
data map $d_\nu({\mathbfit x})$, we define
the continuous isotropic wavelet transform by
\begin{equation}\label{eq:wt}
w_d(R,{\mathbfit b}) = \int d^2{\mathbfit x}\,
\frac{1}{R^2}\psi\left(\frac{|{\mathbfit x} - {\mathbfit b}|}{R}\right)
\,d_\nu({\mathbfit x}),
\end{equation}
where $w(R,{\mathbfit b})$ is the wavelet coefficient associated with the
scale $R$ at the point ${\mathbfit b}$
(where the wavelet is centred).
The function $\psi(|{\mathbfit x}|)$ is the {\em mother} wavelet, which is
assumed to be isotropic and satisfies the conditions
$$
\begin{array}{lr}
\hspace*{-0.8cm}\int d^2{\mathbfit x}\, \psi(x)  = 0 & 
\mbox{(compensation),}\vspace{0.3cm}\\
\hspace*{-0.8cm}\int d^2{\mathbfit x}\, |\psi(x)|^2  =  1 &
\mbox{(normalisation),}\vspace{0.2cm}\\
\hspace*{-0.8cm}C_{\psi} =  
(2\pi)^2 \int_{0}^{\infty} dk\, k^{-1} \,{|\widetilde{\psi}(k)|}^2 < 
\infty &
\mbox{(admissibility)},
\end{array}
$$
where $\widetilde{\psi}$ denotes the Fourier transform of 
$\psi$, $x=|{\mathbfit x}|$ and $k = |{\mathbfit k}|$. The wavelet coefficients 
given by (\ref{eq:wt}) characterise
the contribution from structure on a scale $R$ to the value of the 
map at the position ${\mathbfit b}$.

By analogy with (\ref{reald}), in wavelet space we define the detection
level for the $i$th point source (as a function of scale $R$) by
\begin{equation}
D_w(R) = \frac{w_\xi(R,{\mathbfit b}_i)}{\sigma_{w_n}(R)},
\end{equation}
where $w_\xi(R,{\mathbfit b}_i)$ is the wavelet coefficient 
of the field $\xi_\nu({\mathbfit x})$ at the location 
of the $i$th point source, and 
$\sigma^2_{w_n}(R)$  is the dispersion of the 
wavelet coefficients $w_n(R,{\mathbfit b})$ of the generalised noise field
$n_\nu({\mathbfit x})$. It is straightforward to show that
\begin{equation}
\sigma^2_{w_n}(R) = 2\pi \int_{k_{min}}^{k_{max}} dk\,
k P(k) |\widetilde{\psi}(Rk)|^2,
\end{equation}
where $P(k)$ is the power spectrum of $n_\nu({\mathbfit x})$. 
The integral limits, $k_{min}$ and $k_{max}$, correspond
to the maximum and minimum scales of the sky patch analysed, i.e.
the patch and pixel scales respectively.

%In practice,
%$P(k)$ is approximated by the power spectrum of the full data map
%$d_\nu({\mathbfit x})$, since it is assume that the power spectrum of
%the point source emission $\xi_\nu({\mathbfit x})$ is negligible as
%compared to that of the generalised noise field.

The detection level $D_w(R)$ in wavelet space will have a maximum
value at some scale $R=R_0$. This scale is practically the same for 
all point sources, and  may be found by solving
$dD_w(R_0)/dR = 0$. In general, the optimal scale is of the order of
the beam dispersion $\sigma_a$ (see V01, section 3, for a
discussion about how the noise and the coherence scale of the
background determine the optimal scale).
In order that the wavelet 
coefficients are optimally sensitive to the presence of the point source, 
we must make the value of $D_w(R_0)$ 
as large as possible. This is
achieved through an appropriate choice both of the mother wavelet
$\psi(x)$ and the optimal scale. If the beam profile is Gaussian 
and the power spectrum of the generalised noise field
is scale-free, Sanz et al. (2000) show that for a wide range of
spectral indices of the power spectrum
the Mexican Hat wavelet is optimal.
The two-dimensional MHW is given by
\begin{equation}\label{eq:MHW}
\psi(x) = \frac{1}{\sqrt{2\pi}}\Big[2-\big(\frac{x}{R}\big)^2\Big]
             e^{-\frac{x^2}{2R^2}},
\end{equation}
from which we find
\begin{equation}\label{eq:Coeff}
w_\xi(R,{\mathbfit b}_i) = 2\sqrt{2\pi}\frac{A_i}{\Omega}
                              \frac{(R/\sigma_a)^2}
                              {(1 + (R/\sigma_a)^2)^2},
\end{equation}
where $A_i$ is the amplitude of the $i$th point source, 
$\Omega$ is the area under
the Gaussian beam and $\sigma_a$ is the beam dispersion. 
In (\ref{eq:Coeff}) it is assumed that any overlap of the (convolved)
point sources is negligible. That is a good approximation
for the brightest point sources, the ones that the MHW is able to
detect. 
In fact, the number of point sources detected at each frequency
(see Table~\ref{tps}) corresponds only to a small percentage of the number of
resolution elements (see Table~\ref{observ}) contained at
each Planck frequency channel
($\sim 6\%$ at 30GHz, the most unfavourable case). Therefore,
the probability of finding two or more bright point sources inside the
same resolution element is very low.
The advantage of identifying point sources in wavelet space rather
than real space may then be
characterised by the amplification factor 
\begin{equation}
{\cal A} = \frac{D_w(R_0)}{D} = 2\sqrt{2\pi}
\frac{(R_0/\sigma_a)^2}{(1 + (R_0/\sigma_a)^2)^2}
\frac{\sigma_n}{\sigma_{w_n}(R_0)}.
\end{equation}

In practice, it is clear that we do not have access to the
wavelet coefficients of the fields $\xi_\nu({\mathbfit x})$ and
$n_\nu({\mathbfit x})$ separately, but only to the wavelet coefficients
of the total data map $d_\nu({\mathbfit x})=\xi_\nu({\mathbfit x})+
n_\nu({\mathbfit x})$. Nevertheless, if the detection level
$D_w(R)$ for the $i$th point source is reasonably large, we would
expect $w_\xi(R,{\mathbfit b}_i) \approx w_d(R,{\mathbfit b}_i)$. Also,
if we assume that
the power spectrum of
the point source emission is negligible as
compared to that of the generalised noise field, then
$\sigma_{w_n}(R) \approx \sigma_{w_d}(R)$. Thus our algorithm for
detecting point sources is as follows. Using the above approximations,
we first calculate the optimal scale $R_0$. We then
calculate the wavelet transform
$w_d(R_0,{\mathbfit b})$ of the data map at 
the optimal scale.
The wavelet coefficients are then analysed to find sets of connected
pixels above a certain threshold $\sigma_{w_d}(R_0)$.
The maxima of these
spots are taken to correspond to the locations of the point sources.

%The optimum scale $R_0$ is determined by a 
%combination of the beam size and the characteristics of the
%generalised noise field, but typically $R_0$ is on the order of
%the beam dispersion $\sigma_a$. 
%For data maps
%in which the emission from physical components other than point sources
%has a coherence scale
%larger than the beam dispersion, the optimal
%scale tends to be greater than $\sigma_a$ and the point source
%detection is limited mainly by instrumental noise.
%In our Planck simulations,
%this occurs in the high-frequency channels, where the FWHM 
%of the beam is smaller than the coherence length of the diffuse emission.
%(dust and CMB).
%On the other hand, if the coherence scale of the diffuse emission
%is similar to or lower than the beam dispersion, the optimal
%scale $R_0$ tends to be lower than $\sigma_a$ and
%point source detection is limited by the emission from other
%hysical components. This occurs in the CMB-dominated Planck maps,
%ince the antenna FWHM is similar or greater than
%the CMB coherence scale ($\sim 10'$).

%The MHW filtered, allows to diminish the backfroung influence,
%whereas the noise remains practically indetical.
%The optimal scale determination gets the better point source
%detection for the noise level and the background of the image
%to be analyzed.

For every point source detected in the above way, we then 
go on to estimate its flux. This is achieved by performing a
multiscale fit as follows. For each point source location
${\mathbfit b}_i$, the wavelet transform $w_d(R,{\mathbfit b}_i)$
is calculated at a number of scales $R$ and compared with the
theoretical curve (\ref{eq:Coeff}). This comparison is performed
by calculating the standard misfit statistic
\begin{equation}
\chi^2 =
\left[{\mathbfit w}^{\rm (exp)}-{\mathbfit w}^{\rm (theo)}\right]^{\rm T}
{\mathbfss V}^{-1}
\left[{\mathbfit w}^{\rm (exp)}-{\mathbfit w}^{\rm (theo)}\right],
\end{equation}
where the $k$th element of the vector ${\mathbfit w}^{\rm (exp)}$ is 
$w_d^{\rm (exp)}(R_k,{\mathbfit b}_i)$ (and similarly 
for the vector of theoretical
wavelet coefficients). The matrix ${\mathbfss V}$ is the empirical
covariance matrix of the wavelet coefficients on different scales,
which is given by
\begin{equation}
V_{jk} = \langle w_d^{\rm (exp)}(R_j,{\mathbfit b}) w_d^{\rm (exp)}(R_k,{\mathbfit b})
\rangle,
\end{equation}
where the average is over position ${\mathbfit b}$.

\subsection{MEM and MHW joint analysis}
\label{jointanal}

In H99 the MEM technique is shown to be effective at performing 
a full component separation in the presence of point sources.
In particular, the reconstructed maps of the separate diffuse 
components contain far less point-source contamination than the input
data maps. Moreover, by comparing the true data maps
with simulated data maps produced from the separated components, it is
possible to obtain point source catalogues at each observing frequency.
Nevertheless, the MEM approach does have its limitations. Since the
point sources are modelled as an additional generalised noise
component, it is not surprising that MEM performs well in identifying
and removing the large number of point sources with low to
intermediate fluxes. However, it is rather poorer at removing the contributions
from the brightest point sources. These tend to remain in the
reconstructed maps of the separate diffuse components, although with
much reduced amplitudes.

The MHW technique, on the other hand, performs best when identifying
and removing the brightest point sources. Indeed, in detecting bright
sources the MHW technique generally out-performs other 
techniques such as SExtractor (Bertin \& Arnouts, 1996)
and standard harmonic filtering
(TOC98). Moreover, the amplitudes of the bright sources are also
accurately estimated. For weaker sources, however, the MHW technique
performs more poorly by either inaccurately estimating the flux or
failing to detect the source altogether.

The strengths and weaknesses of the MEM and MHW approaches clearly
indicate that they are complementary techniques, and that a combined
approach might lead to improved results as compared to using each
method independently. We thus propose the following method for
analysing multifrequency observations of the CMB that contain point
source contamination. First, the data map at each observing frequency
is analysed separately using the MHW in order to detect and remove as
many bright point sources as possible and obtain accurate estimates
of their fluxes. The processed data maps are
then taken as the inputs to the generalised MEM approach discussed in
H99. As we will demonstrate in section \ref{foresep}, the MEM analysis 
of these processed maps leads to more accurate reconstructed maps of the
separate diffuse components. This leads in turn to more accurate
residual maps between the true input data with the data simulated
from the reconstructions. These residual maps are then analysed
with the MHW in order both to refine the original estimates of the
fluxes of the bright point sources and to detect fainter
sources. Thus, the joint analysis not only gives a more complete
point source catalogue, but also 
improves the quality of the reconstructed maps of the CMB and other foreground
components.

\section{Simulated Observations}
\label{simobs}

\begin{table}
\caption{Basic observational parameters of the 10 frequency channels
of the Planck Surveyor satellite. Column two lists the
fractional banwidths. The FWHM in column three 
assumes a Gaussian beam.
In column four the instrumental noise level
is $\Delta T$ ($\mu$K) per resolution element for 12 months of
observation.}
\label{observ}
\begin{center}
         \begin{tabular}{|c|c|c|c|}
	 \hline
	 Frequency & Fractional bandwidth & FWHM  & $\sigma_{\rm noise}$ \\
	 (GHz) & ($\Delta \nu /\nu$) & (arcmin) & ($\mu$K) \\
	 \hline
	 30 & 0.20 & 33.0 & 4.4 \\
	 44 & 0.20 & 23.0 & 6.5 \\
	 70 & 0.20 & 14.0 & 9.8 \\
	 100 (LFI) & 0.20 & 10.0 & 11.7 \\
	 100 (HFI) & 0.25 & 10.7 & 4.6 \\
	 143 & 0.25 & 8.0 & 5.5 \\
	 217 & 0.25 & 5.5 & 11.7 \\
	 353 & 0.25 & 5.0 & 39.3 \\
	 545 & 0.25 & 5.0 & 400.7 \\
	 857 & 0.25 & 5.0 & 18182 \\
	 \hline
      \end{tabular}
    \end{center}
\end{table}

As mentioned in the introduction, the joint analysis technique can be
applied to any multifrequency observations of the CMB that may contain
point source emission. Thus, for example, the method could
straightforwardly be applied to the existing Boomerang or MAXIMA
data-sets, or to observations by the forthcoming NASA MAP satellite.
Nevertheless, in order to test the capabilities of the joint analysis
method to the fullest extent, in this paper we apply it to
simulated observations by the proposed ESA Planck Surveyor satellite. 
The basic observational parameters of the Planck mission instruments
(HFI and LFI) are listed in
Table~\ref{observ}.

\begin{figure*}
\resizebox{13cm}{!}
{\includegraphics{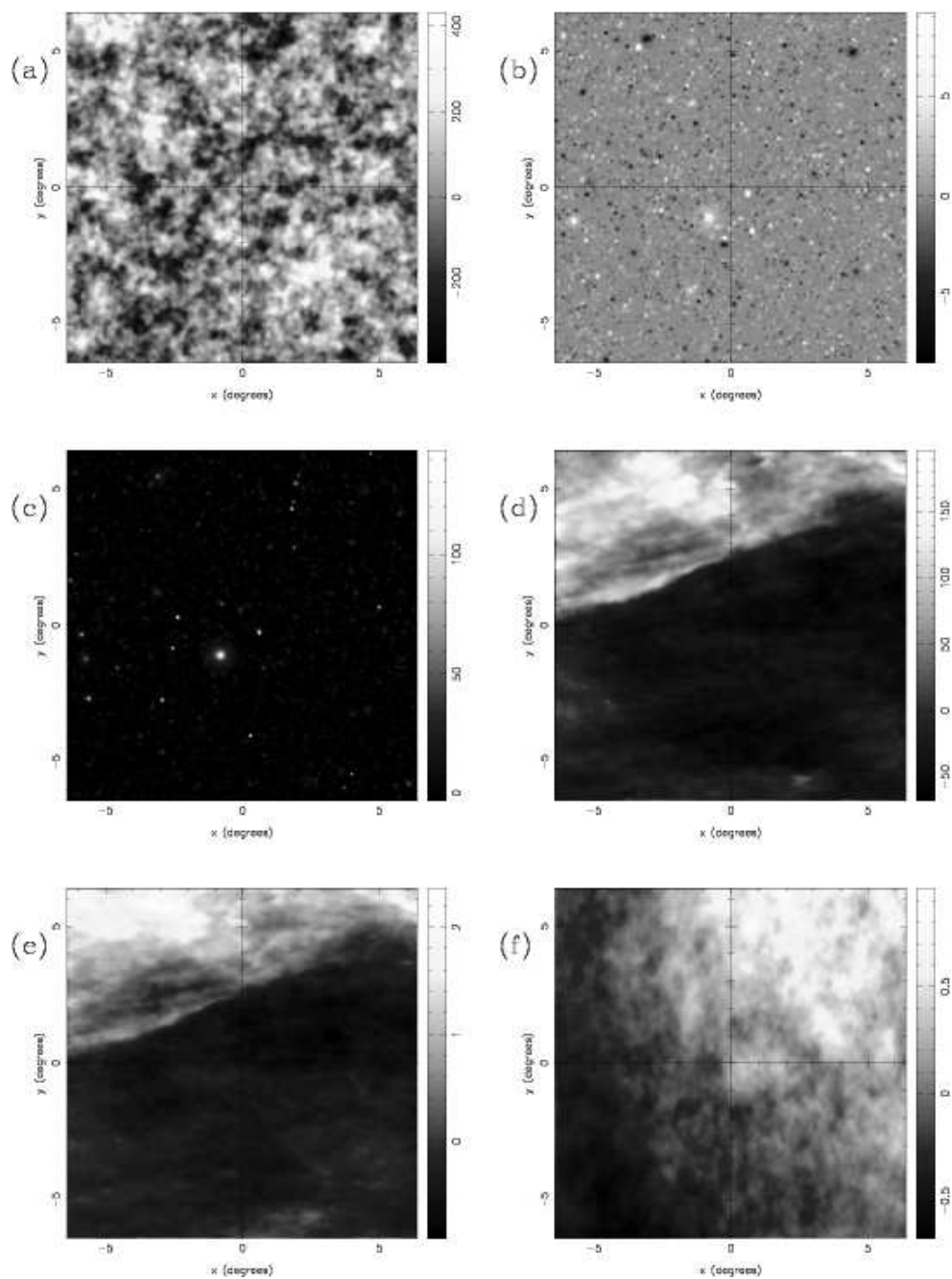}}
\caption[]{The $12.8\times 12.8$ deg$^2$ realisations of the six input
components used to produce the simulated Planck data. The different
panels correspond to (a) CMB,
(b) kinetic SZ effect, (c) thermal SZ effect, (d) Galactic dust,
(e) Galactic free-free and (f) Galactic synchrotron emission. 
Each component is plotted at 300 GHz and has been convolved with a
Gaussian beam of FWHM 5 arcmin (the highest resolution expected for the Planck
satellite). The map units are equivalent thermodynamic temperature in $\mu$K.}
\label{inputs}
\end{figure*}

The simulated data are similar to those used in H99. They include
contributions from the primordial CMB, the thermal and kinetic
Sunyaev-Zel'dovich effects, extragalactic point sources and Galactic
thermal dust, free-free and synchrotron emission. Aside from the point
sources, we assume that the emission of each physical component
can be factorised into a spatial template at 300 GHz with
a known frequency dependence.
In Figure~\ref{inputs} we have plotted the six component templates
at 300 GHz. Each map covers a $12.8 \times
12.8$~deg$^2$ patch of sky and has been convolved 
with a Gaussian beam with FWHM
5 arcmin (i.e. the highest Planck resolution).

The CMB map is a Gaussian realisation of a spatially-flat
inflationary/CDM model with $\Omega_m = 0.3$ and $\Omega_{\Lambda}$ =
$0.7$, for which the $C_\ell$ coefficients were generated using
CMBFAST (Seljak \& Zaldarriaga 1996).  The thermal Sunyaev-Zel'dovich
map is taken from the simulations of Diego et al. (2000) and assume
the same cosmological model as that used for the CMB.  The kinetic SZ
field is produced by assuming the line-of-sight cluster velocities are
drawn from a Gaussian distribution with zero mean and rms $400$
km/s. The
extragalactic point source simulations adopt the model of Toffolatti
et al. (1998) and also assume the same cosmological model.

The Galactic thermal dust emission is created
using the template of Finkbeiner, Davis \& Schlegel (1999). 
The frequency dependence of the dust emission is assumed to follow a
grey-body function characterised by a dust temperature of 18 K and an
emissivity $\beta = 2$. The distribution of 
Galactic free-free emission is poorly known. Current
experiments such as the H-$\alpha$ Sky 
Survey\footnote{http://www.swarthmore.edu/Home/News/Astronomy/}
and the WHAM project\footnote{http://www.astro.wisc.edu/wham/}
should soon provide maps of $H_{\alpha}$ emission that could be used as
templates. For the time being, however, we create a free-free
template that is correlated with the dust emission in the manner
proposed by Bouchet, Gispert \& Puget (1996). 
The frequency dependence of the
free-free emission is assumed to vary as $I_{\nu} \propto
{\nu}^{-0.16}$, and is normalised to give an rms
temperature fluctuation of $6.2 \mu K$ at 53 GHz. Finally, 
the synchrotron spatial template has been produced
using the all sky map of Fosalba \& 
Giardino\footnote{ftp://astro.estec.esa.nl/pub/synchrotron}. This map
is an extrapolation of the 408 MHz radio map of Haslam et al.
(1982), from the original $1$ deg resolution to a resolution
of about $5$ arcmin. The additional small-scale structure
is assumed to have a power-law power spectrum 
with an exponent of $-3$.  We have continued this
extrapolation to $1.5$ arcmin
following the same power law.
The frequency dependence is assumed to be $I_{\nu} \propto {\nu}^{-0.9}$
and is normalized to the Haslam 408 MHz map.

\begin{figure}
{\includegraphics[angle=270,width=8cm]
{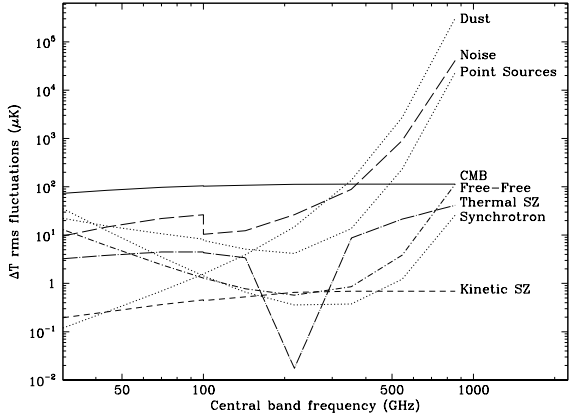}}
\caption[]{The rms thermodynamic temperature
fluctuations at the Planck observing
frequencies due to each physical component, after convolution with the
appropriate beam and asumming a sampling rate of FWHM/2.4. The rms
instrumental noise per pixel at each frequency is also plotted.}
\label{rms}
\end{figure}

\begin{figure*}
\resizebox{13cm}{!}
{\includegraphics{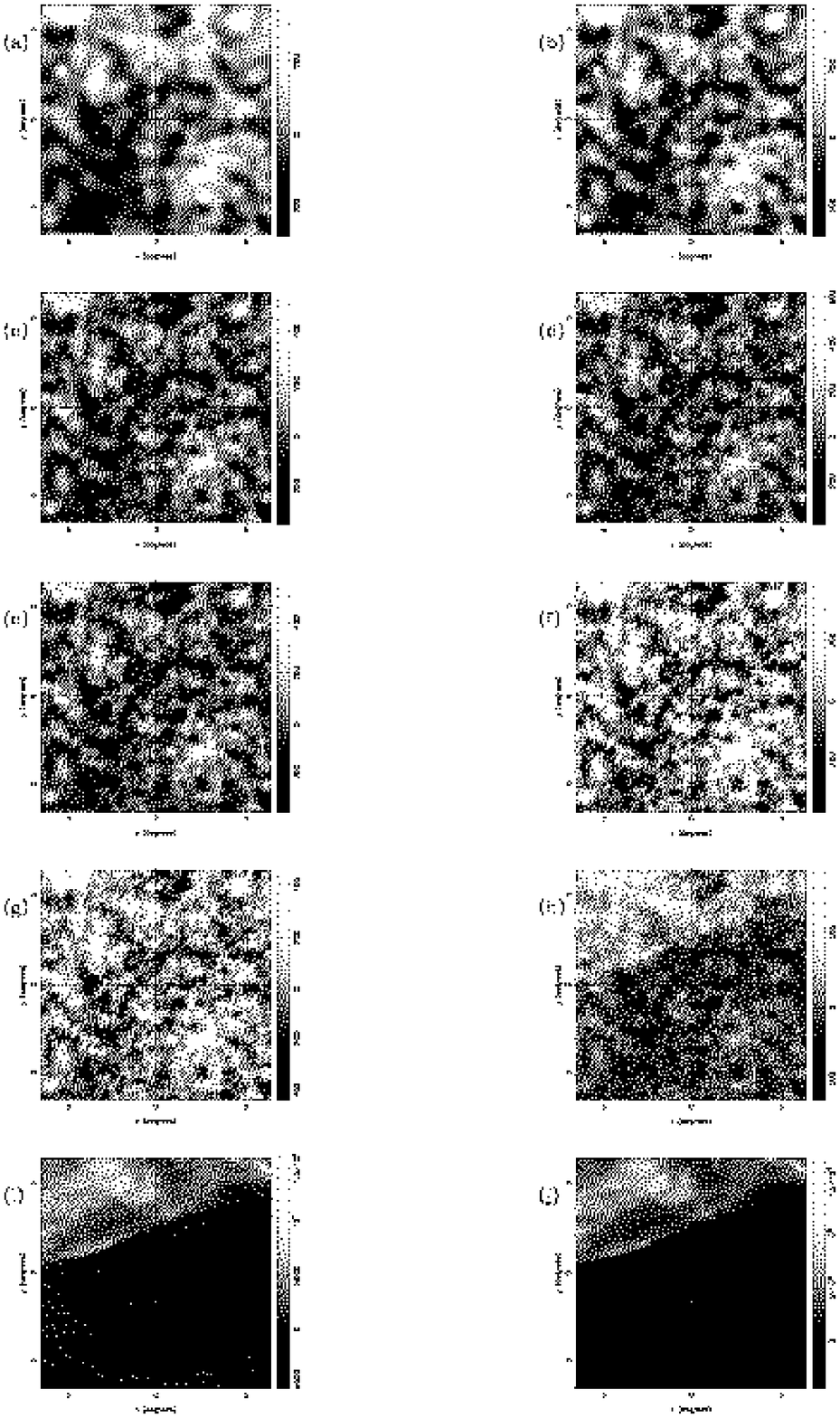}}
\caption[]{The $12.8\times 12.8$ deg$^2$ data maps observed at each of
the ten Planck frequency channels listed in Table~\ref{observ}. The
panels correspond to the frequencies (a)~30~GHz, (b)~44~GHz, 
(c)~70 GHz, (d)~100~GHz-lfi, (e)~100~GHz-hfi, (f)~143~GHz,
(g)~217~GHz, (h)~353~GHz, (i)~545~GHz and (j)~857~GHz. The
units are equivalent thermodynamic temperature in $\mu$K.}
\label{data}
\end{figure*}

To simulate the observed data in a given Planck frequency channel,
each of the physical components discussed above is 
first projected to the relevant frequency and the contributions are
summed. The predicted point source emission for the frequency
is then added, and the resulting total sky emission is 
convolved with a Gaussian
beam of the appropriate FWHM. Finally, independent pixel noise is added,
with corresponding rms from Table~\ref{observ}.
In Figure~\ref{rms} we give the rms thermodynamic temperature fluctuations
in the data at each Planck observing frequency due to each physical
component and the instrumental noise.
In Figure~\ref{data} we plot the simulated Planck observations in each
frequency channel, all the components are included:
CMB, dust, free--free, synchrotron, kinetic and thermal SZ effects and 
point source emission as well as instrumental noise.

\section{Foreground Separation}
\label{foresep}

We have applied the method outlined in section~\ref{jointanal} to
the simulated Planck data described above.  We have assumed knowledge
of the azimuthally averaged power spectra of the six input components
in Fig.~\ref{inputs}, together with the azimuthally averaged cross
power spectra between them (see H98 for more details).  Using the
model of Toffolatti et al. (1998), we have also introduced the power
spectrum of the point sources at each frequency channel, including
cross power spectra between channels, and account for this contaminant
as an extra noise term (see H99 for more details).  However, the
recovery of the main components and point sources does not depend
critically on this assumption, as will be discussed later.

\begin{figure*}
\resizebox{13cm}{!}
{\includegraphics{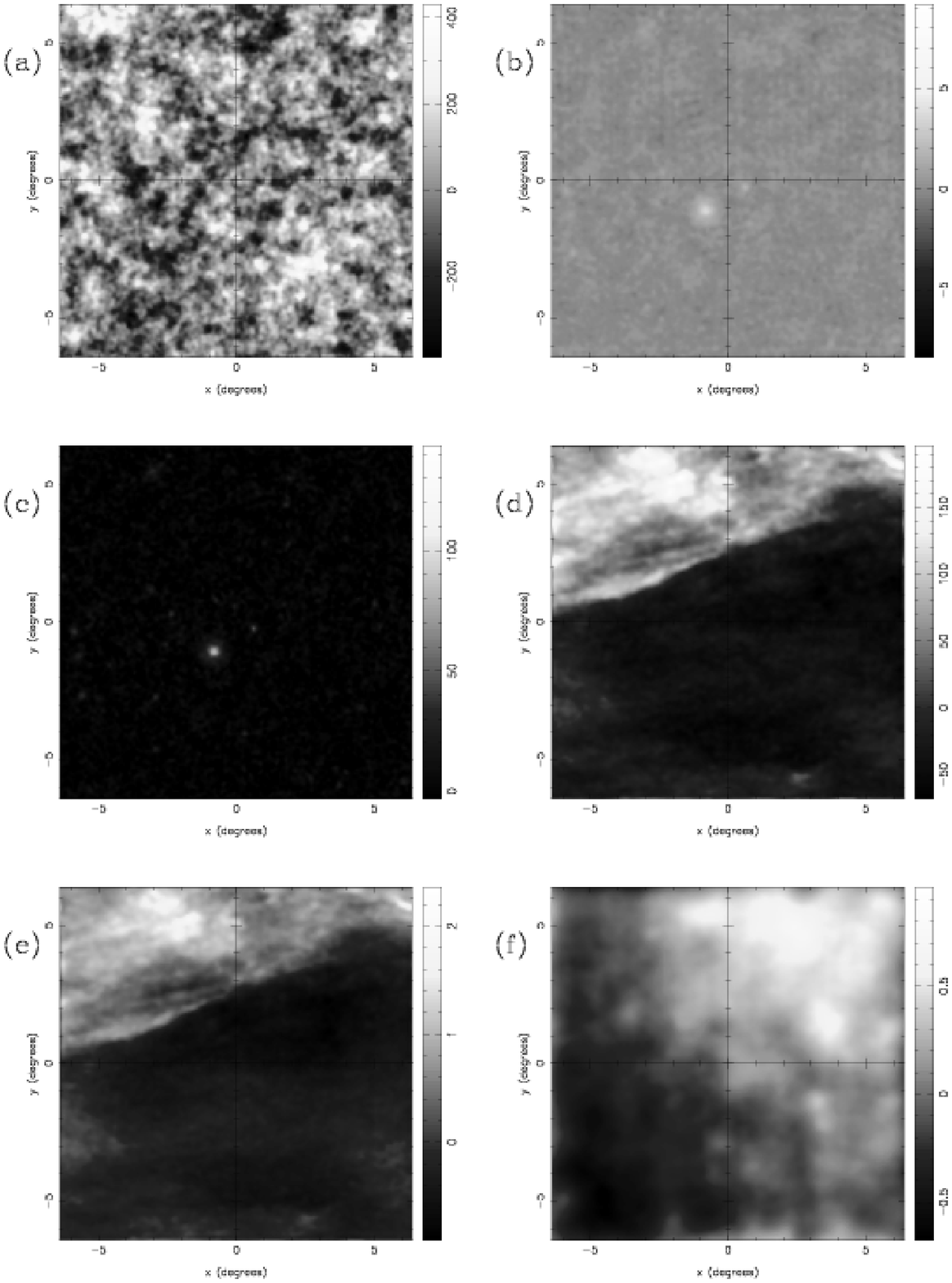}}
\caption[]{The reconstructed maps for each of the physical components:
(a) CMB, (b) kinetic SZ effect, (c) thermal SZ effect, (d) Galactic dust,
(e) Galactic free-free and (f) Galactic synchrotron emission. 
Point sources have been subtracted from the data maps using the
mexican hat algorithm before applying MEM. 
Each component is plotted at 300 GHz and has been
convolved with a Gaussian beam of FWHM 5 arcmin.
The map units are equivalent thermodynamic temperature in $\mu$K.}
\label{rec}
\end{figure*}

\begin{figure*}
\resizebox{13cm}{!}
{\includegraphics{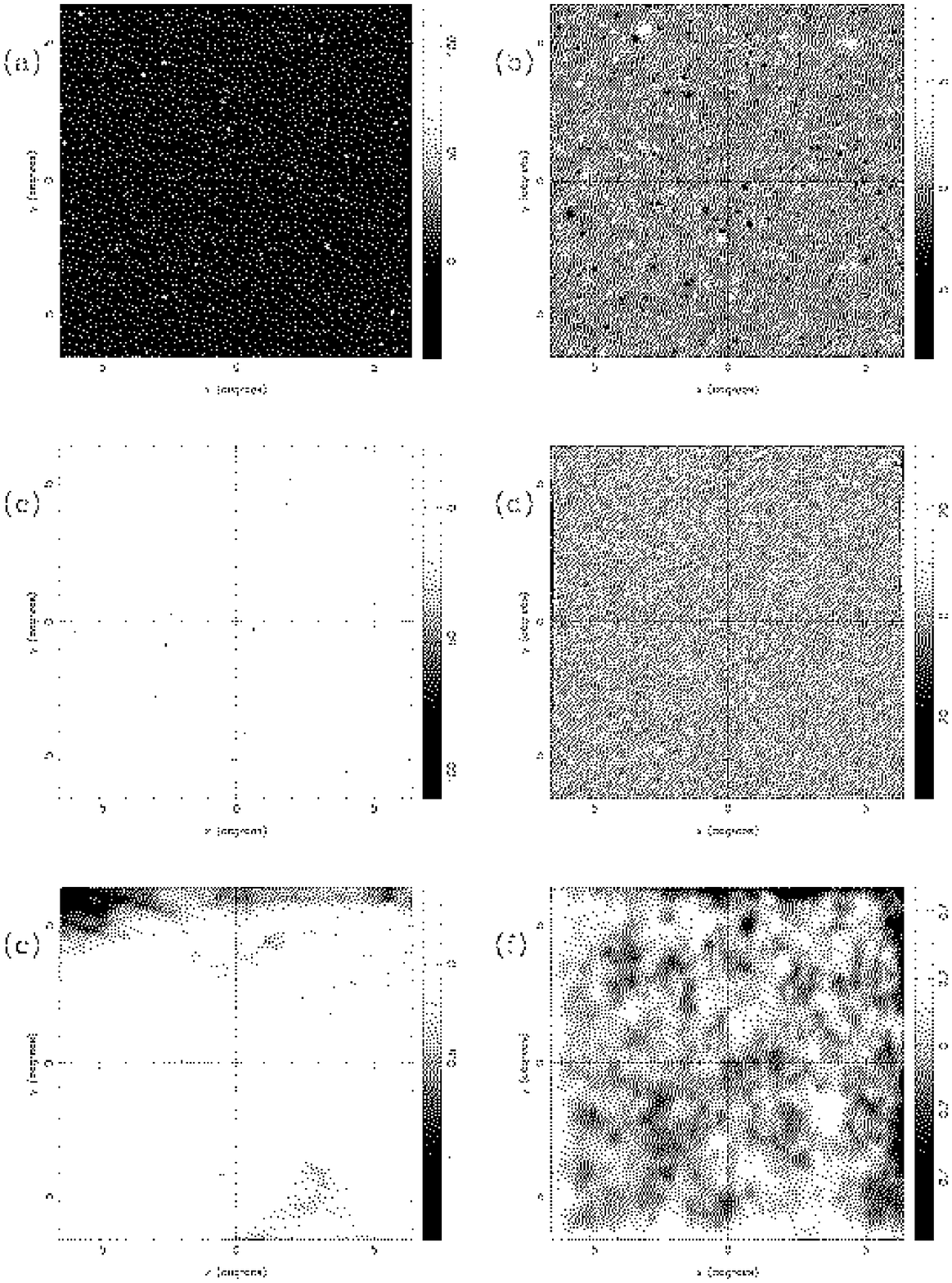}}
\caption[]{The reconstruction residuals obtained from subtracting the
input maps of Fig.~\ref{inputs} from the reconstructed maps of
Fig.~\ref{rec}. The panels correspond to (a) CMB; 
(b) kinetic SZ effect; (c) thermal SZ effect; (d)
Galactic dust; (e) Galactic free-free; (f) Galactic syncrotron
emission.}
\label{res_full}
\end{figure*}

The resulting reconstructions of the physical components at a
reference frequency of 300 GHz are shown in
Fig.~\ref{rec}. The maps have been plotted using the same
grey-scale as in Fig.~\ref{inputs} to allow a straightforward comparison.
In Fig.~\ref{res_full}, we plot the residuals for each component,
obtained by subtracting the input maps from the reconstructions.

We can see that the main input components have
been faithfully recovered with no obvious visible contamination from
point sources. This is because
the MHW subtraction algorithm is efficient at
removing the brightest point sources, whereas MEM has greatly reduced
the contamination due to fainter sources.
We give the rms reconstruction errors for each component in 
Table~\ref{errors}. 

%\textbf{Since these are absolute errors, for
%other frequencies these numbers have to be rescaled following
%the spectral dependence of each component (obviously, for
%the CMB and the Kinetic SZ they remain unchanged).}

\begin{table}
\caption[]{The rms in $\mu$K of the reconstruction residuals smoothed 
with a 5 arcmin FWHM Gaussian
beam with and without the initial
subtraction of bright point sources using the MHW. Full power
spectrum information has been assumed. For comparison the rms of the
input maps shown in Fig.~\ref{inputs} are also given.
Results are given for the reference frequency of 300 GHz.}
\label{errors}
\begin{center}
\begin{tabular}{|c|c|c|c|}
\hline
Component  & input & error & error \\
& rms & (with MHW) & (without MHW) \\
\hline
CMB        & 112.3 & 7.68 & 8.62 \\
Kinetic SZ & 0.69 & 0.70 & 0.70 \\
Thermal SZ & 5.37 & 4.64 & 4.66 \\
Dust       & 55.8 & 2.68 & 3.39 \\
Free-Free  & 0.66 & 0.22 & 0.24 \\
Synchrotron& 0.32 & 0.11 & 0.12 \\
\hline
\end{tabular}
\end{center}
\end{table}

In particular, we note that the CMB has been recovered very
accurately, although the residuals map does show some weak contamination due
to low-amplitude point sources. Indeed, the rms reconstruction error for this
component is $\sim$ 7.7 $\mu$K, which corresponds to an accuracy
of $\sim$ 6.8 per cent as compared to the rms of the input CMB
map (see Table~\ref{errors}).
Even more impressive is the reconstruction of the dust map. None of
the numerous point sources present in the highest frequency channel
maps are visible in the reconstruction. This is also confirmed by
inspecting the residual map. 
The main features of the free-free emission are also
recovered, mostly due to its high correlation with the dust. Again,
the reconstruction shows no evidence of point source contamination.
For the synchrotron component, the recovered emission is
basically a lower resolution image of the input map. This is expected
since only the lowest frequency channels provide useful
information about this component, and these channels also have
the lowest angular resolutions. Although the reconstructed synchrotron map
is mostly free of point sources, some residual contamination remains.
This contamination corresponds to a few medium amplitude point
sources that are present in the lowest frequency channels,
although they are not clearly visible in the data.
These sources are too weak to be detected using
the MHW algorithm but at the same time they are not well
characterised by the generalised noise approach assumed in MEM.

As pointed out in H99, one must be careful when comparing the
amplitude of the residual point sources still contaminating the
reconstructions with the corresponding amplitudes of the point sources
in the data maps. The reconstructions are calculated at a reference
frequency of 300 GHz and those sources remaining in the residuals maps
are projected in frequency according to the spectral dependence of the
component they contaminate.
In addition, we have to take into account the different resolution of
the Planck frequency channels. For example, the 
contaminating point source in the middle right-hand side of the
synchrotron residuals map has an amplitude of $\simeq 0.15\mu$K after
convolution with a Gaussian beam of FWHM 33 arcmin (the
resolution of the 30 GHz Planck frequency channel).
Following the spectral dependence of the synchrotron component, this
projects to $17.5\mu$K at 30GHz. This value should be compared with 
the amplitude of the point source at the same frequency,
which is around $150\mu$K.
%However, this point source in fact has an amplitude around
%150$\mu$K in this frequency channel, which would project to
%1.3$\mu$K at 300 GHz following the spectral dependence of the
%synchrotron component. 
Therefore, MEM has succeeded in reducing 
the contamination due to this point source by almost a factor of ten.

The recovery of the thermal SZ effect is quite good. Most of the
bright clusters have been reproduced whereas only a few point sources
has been misidentified as clusters. At the reference frequency of
300GHz, these misidentified point sources appear mostly as negative
features. Finally, as expected, the reconstruction of the kinetic SZ
is quite poor and one detects only a few clusters whose corresponding 
thermal SZ effect is large. 

We have also calculated the power spectrum of the reconstructed component 
maps and found that the accuracy is very similar to that found in H99,
so we do not plot them again here.
The effect of first applying the MHW to the data maps before the MEM
analysis is not so obvious when considering the power spectra of the 
reconstructions, since
only a small percentage of pixels are affected by residual
point sources and this has little effect in the recovered
spectrum. Nevertheless, the removal of the point source contamination
is vital if one wishes to probe the Gaussian character of the CMB, 
as well as to study properties of the other foreground components.

\begin{figure*}
\resizebox{13cm}{!}
{\includegraphics{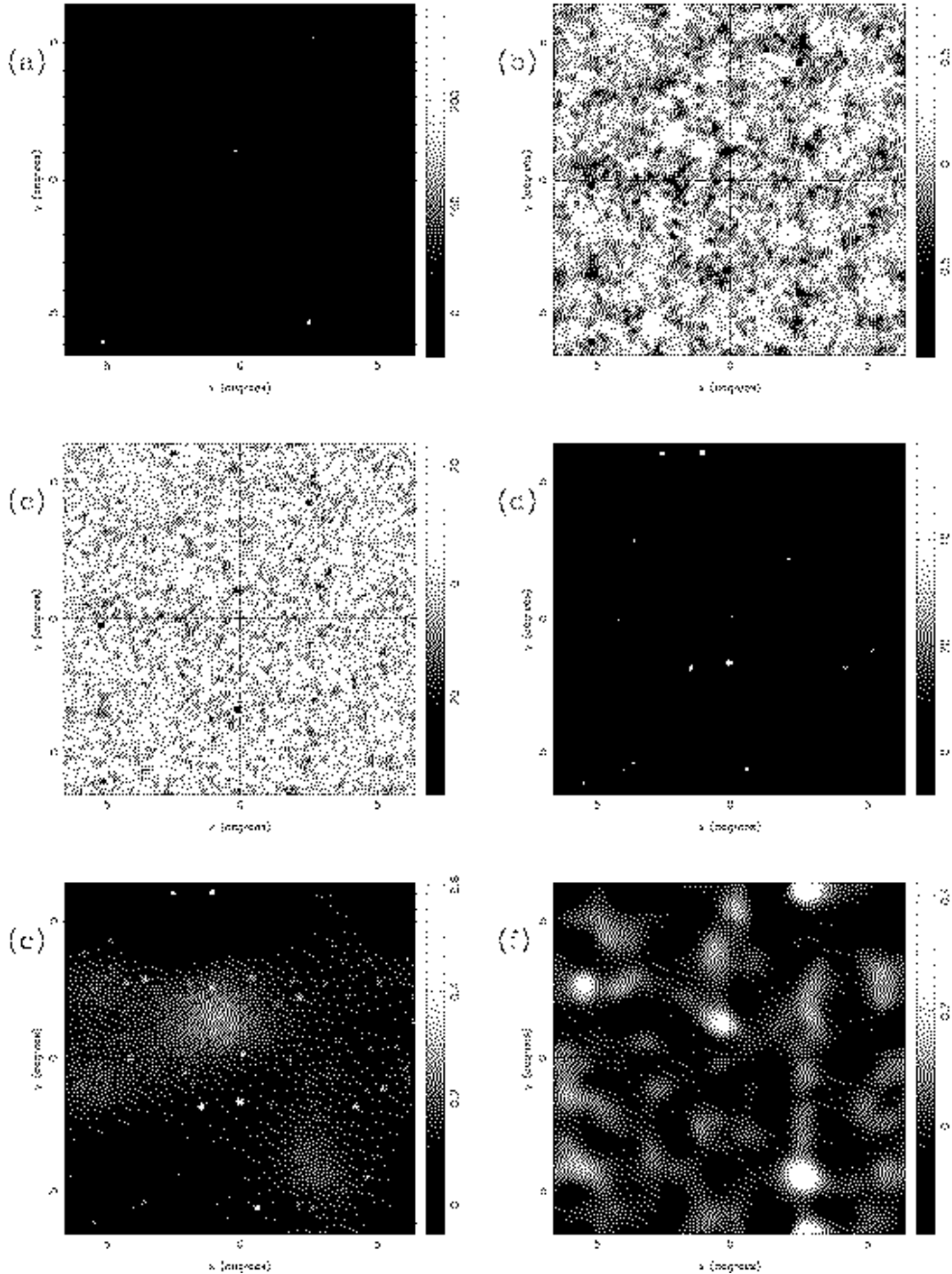}}
\caption[]{The residuals obtained by
subtracting the reconstructed maps
shown in Fig.~\ref{rec} from the reconstructions achieved when 
the MHW is not used to perform an initial point source removal.
Both reconstructions were smoothed
with a Gaussian beam of FWHM 5 arcmin and correspond to a frequency of
300 GHz. The different panels are:
(a) CMB, (b) kinetic SZ effect, (c) thermal SZ effect, (d) Galactic dust,
(e) Galactic free-free and (f) Galactic synchrotron emission. 
The map units are equivalent thermodynamic temperature in
$\mu$K.}
\label{dif}
\end{figure*}

To understand better the effect of performing the MHW analysis on the
data maps prior to the MEM algorithm, we have also carried out a
component separation for the case where the MHW step is {\em not}
performed; this corresponds to the method in H99.
In Fig.~\ref{dif}, we show the difference between the reconstructions
obtained using the combined MHW and MEM technique and those obtained
using MEM alone. Thus these maps display the point sources that have been
successfully removed by the MHW, which would be otherwise
present in the reconstructions. By comparing with the data, we can
also see how the point sources present in the different frequency
channels would affect a given component if not carefully subtracted.
Particularly impressive is the
removal from the dust 
and free-free reconstructions of the large
number of point sources that were present in high-frequency data
channels. For the CMB, the MHW has subtracted a few very bright point 
sources, which dominated the contribution of this contaminant in
the intermediate frequency channels. We also note that the MHW has 
removed a few from the synchrotron reconstruction that were present
in the lowest channels.
The reconstructions of the thermal and kinetic SZ effects have also
been improved since a lower number of point sources have been
misidentified as clusters; these sources were detected by the MHW
mainly in the highest channels of the LFI.
The rms reconstructions errors when MEM is used without a previous
subtraction of point sources by the MHW are also given in 
Table~\ref{errors}.

Finally, we can also study how our reconstructions are affected if we
do not assume full power spectrum information. Thus, we have repeated
our joint analysis of the simulated Planck observations for the 
case where we assume that $\ell^2 C_\ell$ is constant
for each component out to the highest measured Fourier 
mode. The level of the flat power spectrum for each component is, however,
chosen so that the total power in each component is approximately 
that observed in the input maps in Fig.~\ref{inputs}. 
Furthermore, no information about the cross power spectra between
different components is given. 
Regarding the point sources, the true azimuthally averaged
power spectrum is again assumed to account for their contribution
as an extra noise term but cross-correlations between different
frequency channels are ignored.
%The reconstructed maps for this case are given in Fig.~\ref{rec_noinf}
%and the correspoding rms errors are included in Table~\ref{errors}.
The quality of the reconstructions of the main components is actually
very similar to the case when full power spectrum information is
given. In particular, the accuracy of the
CMB reconstructed map is only slightly worse with a rms 
error of 8.2$\mu$K as compared to 7.7$\mu$K in the former case.
Moreover, the reconstruction is again free from
obvious contamination due to point sources.
Similarly, the dust component has been faithfully recovered with a rms
error of 3.0$\mu$K versus 2.7$\mu$K when full power spectrum is
assumed. The main features of the synchrotron and thermal SZ effect
are also recovered although the reconstructions are poorer. In
particular, the contamination due to point sources is slightly
increased in the thermal SZ reconstructed map. Finally, the weakest
components, free-free emission and kinetic SZ effect, are lost in this
case and the
reconstructions have simply defaulted to zero in the absence of any
useful information.

\section{Recovery of point sources}
\label{ps}

\begin{table*}
\begin{minipage}{170mm}
\caption[]{The point source catalogues obtained using the MHW alone
(MHWc), MEM alone (MEMc) and the joint analysis method (M\&Mc).
For each Planck observing frequency, we list the number of sources
detected, the flux limit of the catalogue and 
the average of the absolute value of the percentage error
%the mean percentage absolute error 
for the amplitude estimation. The numbers in brackets in the
second column indicate the number of point sources that
are detected by the $5\sigma_{w_d}$ criterion (see text for details).}

\label{tps}
\begin{center}
\begin{tabular}{|c|c|c|c|c|c|c|c|c|c|}
\hline
& & MHWc & & & MEMc & & & M\&Mc & \\
\hline
Frequency &
Number of & Min Flux & $\overline{E}$ &
Number of & Min Flux & $\overline{E}$ &
Number of & Min Flux & $\overline{E}$ \\
(GHz) &
detections & (Jy) & (\%) &
detections & (Jy) & (\%) &
detections & (Jy) & (\%) \\
\hline

30 & 15 (4)& 0.13 & 14.0 &
	32 & 0.09 & 14.9 &
	30 & 0.09 & 14.0 \\
44 & 11 (3) & 0.27 & 11.2 &
	28 & 0.11 & 14.2 &
	28  & 0.11 & 13.2 \\
70 & 7 (5) & 0.30 & 7.9 &
	38 & 0.08 & 11.8 &
	35 & 0.09 & 11.4 \\
100 (LFI) & 12 (3) & 0.23 & 11.8 &
	37 & 0.08 & 16.4 &
	44 & 0.06 & 18.4 \\
100 (HFI) & 17 (7) & 0.14 & 14.1 &
	72 & 0.04 & 17.3 &
	74 & 0.03 & 17.4 \\
143 & 11 (4) & 0.16 & 18.0 &
	0 & -- & -- &
	5 & 0.32 & 12.4 \\
217 & 15 (5) & 0.08 & 14.4 &
	0 & -- & -- &
	5 & 0.25 & 15.4 \\
353 & 16 (10) & 0.15 & 14.3 &
	38 & 0.08 & 28.7 &
        37 & 0.08 & 34.9 \\
545 & 37 (29) & 0.29 & 14.1 &
	89 & 0.18 & 26.4 &
	121 & 0.15 & 17.2 \\
857 & 306 (86) & 0.30 & 16.5 &
	458 & 0.23 & 20.6 &
	492 & 0.19 & 19.5 \\
\hline
\end{tabular}
\end{center}
\end{minipage}
\end{table*}

The previous section focused on the recovery of the six physical
component maps shown in Fig.~\ref{inputs}. However, a major aim of the
Planck mission is also to compile point source catalogues at each of its
observing frequencies. In this section, we compare the
catalogues obtained using MEM alone (i.e., without a previous
subtraction of bright sources detected by the MHW), MHW alone and
the joint analysis method M\&M.

The MHW catalogue (MHWc) is produced in the manner explained in
V01, through the so called $50\%$ error criterion (see
the previous work for more details in detection criteria).
In short, each of the data maps of Fig.~\ref{data} 
is independently analysed with the MHW. 
Those coefficients above $2\sigma_{w_d}$ at the optimal
wavelet scale are identified as point source candidates. A multiscale
fit is then performed in order to estimate their amplitude and
those wavelet coefficients with a non-acceptable fit are discarded.
We then look for the flux above which at least 95 per cent of the
the detected point sources have a relative error $\le 50\%$.
This gives our estimation of the flux limit. 
Thus, the number of detections at a given channel 
is given by those point sources
with an estimated amplitude above the flux limit.
In practice, we also use multifrequency information to include 
those point sources that, having
an error larger than $50\%$ or an insufficiently good fit,
have been detected in an adjacent channel (although in most channels 
this only accounts
for a very small fraction of the detected point sources).
The error is defined as:
$E = |A - A^{e}|/A$,
where $A$ is the flux of the simulated
source and $A^{e}$ that of the estimated one.

Although the MHW catalogue (hereafter MHWc) is obtained 
in the same way as that of V01, the results here
differ slightly. This apparent discrepancy is due to the different
sampling rates that have been considered in each case. V01 assumes
pixel sizes of 1.5, 3 and 6 arcminutes for the different Planck
channels, whereas in this work the pixel size is given by a fixed
sampling of 2.4, following by a regridding to pixels of size 1.5
arcminutes. Therefore, the number of detected point sources
and fluxes of V01 and this paper are not directly comparable.

Nevertheless, not all the point sources in the MHWc are
subtracted from the original maps prior to performing the
MEM analysis. In theory, giving as much information as available
should improve the MEM results.
However, when using the $50\%$ error criterion a significant 
number of point sources, especially at the 545 and 857 GHz channels,
are estimated with a large error. 
Therefore, if this information
is given to MEM (by subtracting these sources from the original maps),
we are misleading the MEM algorithm. 
To avoid this unwanted effect, the point sources subtracted from 
the maps should be those with the lowest errors in the amplitude estimation.
We need a more robust criterion than that of the $50\%$ error.
Instead we adopt the so called $5\sigma_{w_d}$ criterion which is
also explained in V01. Briefly, we consider
that a point source has been detected if the position
of its maximum is above $5\sigma_{w_d}$ and its
multiscale fit is acceptable.

The MEM catalogue (MEMc) and the M\&M one (M\&Mc)
are obtained using the method outlined in H99.
First, the reconstructed maps of Fig.~\ref{rec} are used as inputs to
produce `mock' data. We follow the same procedure as that used
to obtain the data of Fig.~\ref{data} but, of course, we do not add
instrumental noise or the point sources.
These mock data are then subtracted from the true data (which contain
the full point source contribution). Since the reconstructions
of the six main components are reasonably accurate and also
contain very few point sources, we obtain a data residuals map 
at each Planck frequency that consists mainly of the point sources and
instrumental noise.  Each of these residuals maps are then
independently analysed in order to produce a point souce catalogue at
each observing frequency. We point out, however, that the
residuals maps produced here differ from those in H99.
In order to concentrate on the effect of emission from 
other physical components on the point source recovery,
in H99 the instrumental noise was neglected when making the residuals maps.
Here the instrumental noise is included to obtain a more realistic
estimate of the number of points sources recoverable from real 
Planck data.

Another difference with H99 is the process by which the point source 
catalogue is produced from the residuals maps.
In H99, the SExtractor package (Bertin \& Arnouts, 1996)
is used to detect and estimate the amplitude of point sources. 
The SExtractor package begins by fitting and subtracting 
an unresolved background component, before identifying any 
point sources, and can lead to ambiguities in assigning a flux 
detection limit. Therefore, in the present paper,
the residuals maps are instead analysed using the MHW,
since this wavelet filter is optimal for this purpose
(Sanz et al. 2000).
At this point, however, it is worth pointing out some
subtleties associated with applying the MHW in these circumstances.
In particular, for the MEMc and M\&Mc, 
we apply the $50\%$ error criterion into the residual maps,
in order to compare with the MHWc.
Clearly, this choice determines empirically the
flux limit of the catalogue
(achieved by the $50\%$ error criterion) and
will depend on the assumed point source
population model, but we expect that most models lead to similar
results. In fact, in V01 it is shown that for the Guiderdoni et al.
(1998) E model, the flux limits achieved are very similar.

In Table~\ref{tps} we give, for each catalogue, 
the number of point sources detected, the minimum
flux achieved and the average amplitude error in
each Planck frequency channel.

In the two highest frequency channels,
the joint analysis clearly outperforms the results obtained
with each of the methods separately.
This is due to the complementary nature of the two approaches,
so that bright sources are detected by MHW and intermediate
flux sources are identified by MEM. If MEM alone is used,
many of the brightest point sources
remain in the MEM reconstructions since they are not well
characterised by a generalised noise. Therefore they are either not
detected in 
the data residuals maps or the error in the estimated amplitude is
significantly large. However, in the joint analysis these
sources are easily detected and their fluxes accurately estimated.

Regarding the 353~GHz channel, the number of point sources
detected with M\&M is comparable to the best of the individual
methods, i.e., when using MEM on its own.
This is due to the fact that the main contaminant of the residuals
maps is the high level of noise of the 353~GHz channel, 
which is equally present in both the residuals obtained with MEM and
with M\&M. In fact, many of the point sources have been 
detected (with a large error) thanks to the multifrequency
information.

The low number of point sources detected at intermediate frequencies
(217 and 143 GHz) 
in the M\&Mc in comparation with the MHWc,
can be explained as a combination of factors. First of all, due to the 
high level of noise in these channels relative to the point source
emission, only a few point sources are detected with the MHW in the
first step of our analysis, through the $5\sigma_{w_d}$ criterion.
Therefore, when applying MEM, part of
the emission of the undetected point sources is left in the
reconstructed components, being mainly misidentified with CMB, which
is the dominant component at those frequencies. This has the effect of 
lowering the level of the point sources in the data residuals, which
together with the high level of noise, leads to a low number of
recovered point sources.

In the low-frequency channels, the number of point sources detected
by the joint analysis is similar to that by MEM alone
(the fact that M\&M does not work better than MEM alone for all the
channels may be understood as a statistical fluctuation;
we expect that with an all sky point source simulation, the results of the
combined method will be better than those of MEM in every case). In this case,
the MHW contribution is to improve the amplitude estimation of 
a few point sources, which leads to a lower average amplitude error.
In these frequency channels MHW alone detects only the few
brightest point sources. This is because these channels are dominated 
by the CMB and the beam FWHM is so large that the CMB and
point sources have a similar characteristic scale.

Regarding the estimation of point source amplitudes, the joint
analysis also performs better than each method independently.
Although the average error in the amplitude estimation 
can be larger in the M\&Mc than in the MHWc 
due to the detection of a larger number of
faint point sources, those point sources present in all three
catalogues are, on average, better estimated with the combined
analysis.

\begin{figure*}
\resizebox{13cm}{!}
{\includegraphics{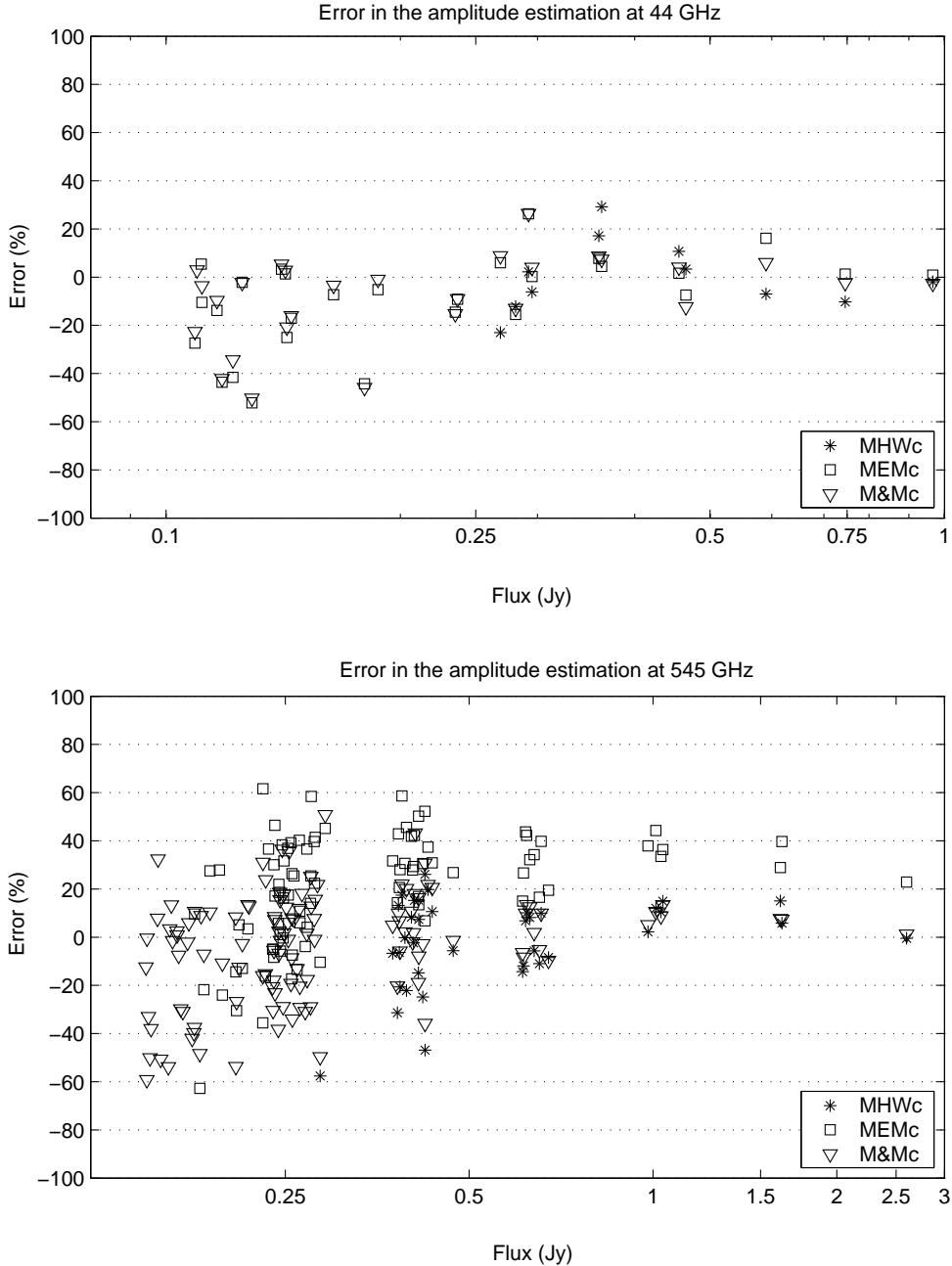}}
\caption{Errors in the amplitude estimation for the MHW, MEM and joint
analysis catalogues (MHWc, MEMc and M\&Mc, respectively)
as a function of the flux. We plot
two Planck frequencies: 44 GHz (top) and 545 GHz (bottom).}
\label{cat_errors}
\end{figure*}

In Figure~\ref{cat_errors} we plot the amplitude estimation errors for
the MHWc, MEMc and M\&Mc versus the true flux (in Jy) for two 
representative channels: 44 and 545 GHz. 
We can see that there is a clear bias in the estimation of the
amplitude of the
brightest point sources in MEMc since they remain in the
reconstructions and are therefore underestimated (corresponding to
positive errors in Fig.~\ref{cat_errors}). 
This problem is solved when combining MEM and the MHW. It is also
obvious that a larger number of point sources and fainter fluxes are
achieved in the combined analysis with respect to the MHW on its own.

\begin{figure*}
\resizebox{18cm}{!}
{\includegraphics{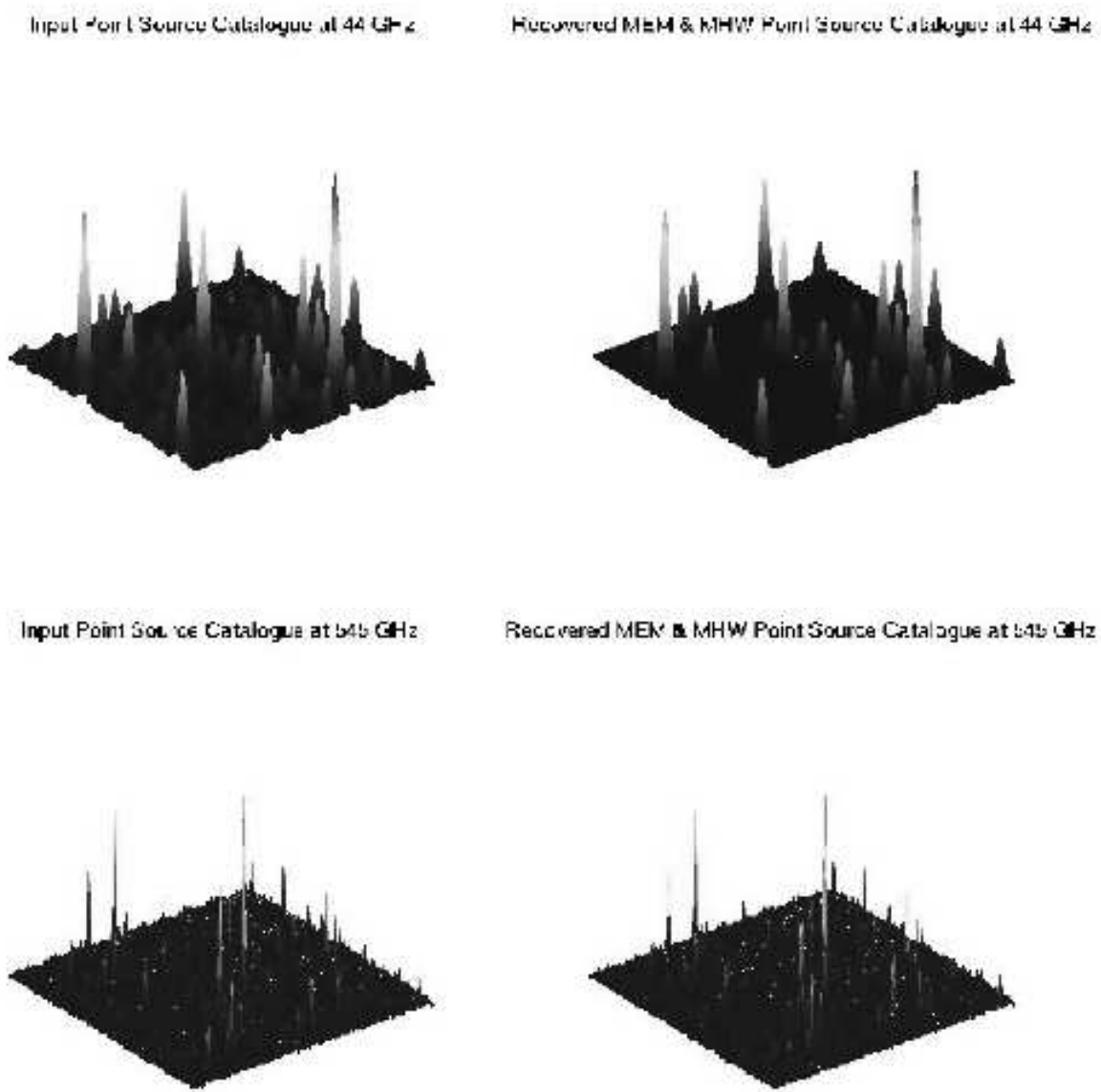}}
\caption{Input and recovered point source catalogues for the
44 GHz (top) and 545 GHz (bottom) Planck channels. The catalogues are convolved
with the corresponding Planck beams}
\label{cat_maps}
\end{figure*}

In Figure~\ref{cat_maps} we have plotted the sources in M\&Mc,
for the same two channels, together with the input point sources
maps. We see that the main features are very well recovered.
At 44 and 545 GHz, the flux limit is
comparable to the level of instrumental noise (see next Section).
Thus, to increase still further the number of point sources detected
and reach fainter fluxes, one would need to denoise the residuals
maps; this is discussed in the next section.

We have also investigated the effect of reducing the
power spectrum information given to MEM. We find that even in the extreme
case when a flat power spectrum is assumed for
the different components,
the results are not significantly different for the M\&Mc, but the
quality of the MEMc is somewhat reduced, especially in the 
high frequency channels. In particular, the amplitude
estimation errors are higher and the catalogue flux limit increases
slightly.

\section{Discussion and conclusions}

\begin{table}
\caption{M\&M estimation of a two-thirds of the sphere
point source catalogue (see text for details).}
\label{23sphere}
\begin{center}
\begin{tabular}{|c|c|c|c|}
\hline
Frequency & $\sim$ counts in & $\sim$ counts in & (\%) \\
(GHz) &  M\&Mc & the model & \\
\hline
30 & 5000 & 7500 & 66  \\
44 & 4500 & 4500 & 75 \\
70 & 5800 & 6800 & 85 \\
100 (LFI) & 7000 & 10800 & 65  \\
100 (HFI) & 12000 & 24300 & 49 \\
143 & 800 & 900 & 89 \\
217 & 800 & 850 & 94 \\
353 & 6000 & 13000 & 46 \\
545 & 20000 & 32000 & 63 \\
857 & 82000 & 200000 & 41 \\
\hline
\end{tabular}
\end{center}
\end{table}

The MEM (H98, H99) and the MHW (C00, V01) techniques have
complementary characteristics when
recovering the microwave sky. On the one hand, the MEM technique is a
powerful tool for 
using multifrequency data to separate the cosmological signal from
foreground emission whose spectral behaviours are (reasonably) well-known.
The most problematic foreground to remove is that due to
point sources. On the other hand,
the MHW has shown to be a robust and self-consistent
method to detect and subtract this point source emission
from microwave maps.
The aim of this paper has been to show how the performance
of a combined (MEM and MHW) analysis can improve the recovery of the
components (CMB, Sunyaev-Zel'dovich, extragalactic
point sources and Galactic emission) of simulated
microwave maps.
In order to test this analysis, we have applied it to
simulated ESA Planck satellite observations. However,
the technique could straightforwardly be applied
to other CMB experiments (e.g. NASA MAP satellite, Boomerang
and MAXIMA).

The proposed method to analyse these data is as follows.
First, we apply the MHW at each observing frequency
in order to remove the brightest point sources
and obtain very good amplitude estimations.
The MEM technique is then applied to these maps to
reconstruct the different components (except the remaining point sources
contribution, which is treated as an additional `noise').
Following the approach discussed in H99, we generate
mock observed data from our reconstructions.
These maps are then subtracted from the initial data.
This provides data residuals maps which mostly contain
instrumental noise plus point source emission
(with slight traces of unrecovered diffuse components).
These residual maps are then analysed again with 
the MHW in order to refine the number of detections and the
amplitude estimation of the point sources.

As already discussed in section \ref{foresep},
the joint analysis improves the accuracy of
the component separation of all the
diffuse components.
This is so because the MHW subtraction algorithm
is efficient at removing the brightest point
sources, whereas MEM has greatly reduced
the contamination due to fainter sources.
We compare the reconstructions achieved when the MHW is or is not
applied. In particular, Figure \ref{dif} shows how
many point sources would remain in the reconstructed
components if the MHW were not used.
We can see that a large number of point sources are removed from the
dust and free-free maps.
There are also a handful of point source that would contaminate 
the synchrotron emission, coming from the low frequency channels.
A lower number of point sources would affect the
CMB reconstruction since the cosmological signal is the main component 
at the intermediate Planck frequencies, where point source
emission is lower. 
Finally, a few point sources would be misidentified with SZ clusters,
appearing in the reconstruction at the reference frequency as sharp
negative features. 

In the previous section, we gave estimates
of the point source catalogues that MEM, MHW and the
joint method provide for these simulations (see Table~\ref{tps}).
We see that the joint analysis provides, in general, a more
complete catalogue than each of the methods on its own, 
reaching lower fluxes and with point source amplitude more accurately
estimated. The improvement is especially
clear at high frequencies due to the high resolution of those Planck
channels. The differences between the number of detections in the MEMc
and M\&Mc are smaller for the channels between 30 and 100 GHz.
This is due to the difficulty of detecting point sources
using the MHW when the background has a similar scale variation to
that of the point sources (see V01 for more details). Hence the main
contribution of the MHW at these frequencies is in improving the amplitude
estimation. In Table \ref{23sphere} we give an estimate
of the number of point sources that would be detected with this combined method
in two--thirds of the sky after 12 months
of observation with the Planck satellite. 
This number is simply obtained by multiplying the counts of Table~\ref{tps}
by the ratio between the solid angle covered by two-thirds of the
sphere and that covered by our simulations.
We compare the recovered point source catalogue with the
simulated one, with a cutoff as given by the
`Minimum Flux' column for M\&M given in Table~\ref{tps}.
We can see that for most
of them, the percentage of detection is around or above $50\%$.
Current evolution models of dust emission in galaxies
(see, e.g., Franceschini et al. 1994; Guiderdoni et al. 1998; Granato et al. 2000) 
give different predictions for counts in the high frequency
Planck channels. On the other hand, all these models predict
a very sharp increase of the far--IR/sub--mm galaxy counts at fluxes
$\simeq 20-100$ mJy. Therefore, given the detection limits of
Table~\ref{tps}, Planck data alone will not be able
to disentangle among different models, although it could 
marginally detect the sharp increase
in the counts in the channels where the minimum flux achieved lies
below 100 mJy. In any case, Planck
will provide very useful data on counts, in a flux range not
probed by other experiments. These data, complementary to the
deeper surveys from the ground or from the space
(ESA FIRST and ASTRO--F/IRIS missions), will surely allow to
discriminate among the various evolutionary scenarios.

Spectral information about the point sources could also be
used to improve further the recovered catalogues. Indeed,
V01 have shown that following point sources through adjacent channels,
one can estimate the spectral indices of the different point source
populations. 
This would allow the recovery of point sources that,
albeit below the detection limit, have an amplitude and
position in agreement with those predicted from adjacent channels.

Finally, as pointed out in the previous section,
the flux limits achieved in the M\&Mc are
close to the noise level. Indeed, the faintest point
sources detected in the catalogue have a fluxes which are
3.0, 2.4, 1.6, 1.2, 1.1, 11.2, 6.7, 1.1, 1.3 and 1.4
times the noise rms in 
the 30, 44, 70, 100(LFI), 100(HFI), 143, 217,
353, 545 and 857 GHz channels, respectively. 
To reach fainter fluxes in these channels is a difficult task,
since we are very close to the noise level except for
the 143 and 217 GHz channels.
On the other hand, if we subtract the MHWc sources from
the original data at 143 and 217 GHz,
instead of the point sources detected by the $5\sigma_{w_d}$
criterion, we could greatly increase the
number of sources and the depth of the M\&Mc at those
frequencies.
A possibility to improve the results
at all frequencies could be 
to denoise these
data residual maps. One way is using wavelet techniques that
have been proved to 
be very efficient at removing noise from CMB maps (Sanz et
al. 1999a,b). However, care must be taken when denoising the
residual maps since the denoising procedure may change the profile of
the point source in wavelet space.
A detailed study of the properties of the denoised map would then
become necessary. In this case, instead of the Mexican Hat, one could
use a customised pseudo-filter to detect point sources in the
residual maps as proposed by Sanz et al (2000).

A natural way to improve the results is to subtract
the recovered M\&Mc from the original data and applying
the MEM algorithm again. This process could be performed iteratively
until the flux limits and the number of counts converge.
However, this method has some disadvantages. As pointed out
in the previous section, if the sources subtracted from the input maps
have a large error, this could mislead the MEM algorithm. This is
the reason why we choose to subtract the catalogue achieved by the
$5\sigma_{w_d}$ criterion instead of subtracting the one given by the
$50\%$ error criterion. The number of point sources with
\emph{large errors} in the M\&Mc is larger than in the MHWc obtained
with the $50\%$ error criterion. 
Hence, a more detailed analysis becomes necessary in order to improve
the results with an iterative approach. Such a study will be performed in a
future work, where the combined technique will be extended to the
sphere. Moreover, the flux limits are already close to the noise
level and thus we do not expect the detection levels to change
substantially (except for the 143 and 217 GHz channel, that can be
clearly improved).

\section*{Acknowledgments}
PV acknowledges support from Universidad de Cantabria fellowship as
well as the Astrophysics Group of the Cavendish Laboratory for their 
hospitality during April 2000.
RBB acknowledges financial support from the PPARC in the form of a
research grant.
PV, EMG, JLS and LT thank Spanish DGESIC Project no.
PB98-0531-c02-01 for partial support. EMG and JLS thank FEDER Project
no. 1FD97-1769-c04-01 and EEC Project INTAS-OPEN-97-1192 for partial
financial support.

\bsp

\label{lastpage}

\end{document}